# hStorage-DB: Heterogeneity-aware Data Management to Exploit the Full Capability of Hybrid Storage Systems

Tian Luo[1]   Rubao Lee[1]   Michael Mesnier[2]   Feng Chen[2]   Xiaodong Zhang[1]

[1]The Ohio State University
Columbus, OH
{luot, liru, zhang}@cse.ohio-state.edu

[2]Intel Labs
Hillsboro, OR
{michael.mesnier, feng.a.chen}@intel.com

## ABSTRACT

As storage systems become increasingly heterogeneous and complex, it adds burdens on DBAs, causing suboptimal performance even after a lot of human efforts have been made. In addition, existing monitoring-based storage management by access pattern detections has difficulties to handle workloads that are highly dynamic and concurrent. To achieve high performance by best utilizing heterogeneous storage devices, we have designed and implemented a heterogeneity-aware software framework for DBMS storage management called hStorage-DB, where semantic information that is critical for storage I/O is identified and passed to the storage manager. According to the collected semantic information, requests are classified into different types. Each type is assigned a proper QoS policy supported by the underlying storage system, so that every request will be served with a suitable storage device. With hStorage-DB, we can well utilize semantic information that cannot be detected through data access monitoring but is particularly important for a hybrid storage system. To show the effectiveness of hStorage-DB, we have implemented a system prototype that consists of an I/O request classification enabled DBMS, and a hybrid storage system that is organized into a two-level caching hierarchy. Our performance evaluation shows that hStorage-DB can automatically make proper decisions for data allocation in different storage devices and make substantial performance improvements in a cost-efficient way.

## 1. INTRODUCTION

Database management systems (DBMSs) have complex interactions with storage systems. Data layouts in storage systems are established with different types of data structures, such as indexes, user tables, temporary data and others. Thus, a DBMS typically issues different types of I/O requests with different quality of service (QoS) requirements [13]. Common practice has treated storage as a black box for a long time. With the development of heterogeneous devices, such as solid-state drives (SSDs), phase-change memories (PCMs), and the traditional hard disk drives (HDDs), storage systems are inevitably becoming hybrid [5, 6, 17, 20]. The "black-box" concept of management for storage is hindering us from benefiting from the rich resources of advanced storage systems.

There are two existing approaches attempting to best utilize heterogeneous storage devices. One is to rely on database administrators (DBAs) to allocate data among different devices, based on their knowledge and experiences. The other is to rely on a management system where certain access patterns are identified by runtime monitoring data accesses at different levels of the storage hierarchy, such as in buffer caches and disks.

The DBA-based approach has the following limitations: (1) It incurs a significant and increasing amount of human efforts. DBAs, as database experts with a comprehensive understanding of various workloads, are also expected to be storage experts [3]. For example, it is a DBA's decision to use certain devices (such as SSDs) for indexes and some frequently used small tables, and less expensive devices (such as HDDs) for other data. (2) Data granularity has become too coarse to gain desired performance. As table size becomes increasingly large, different access patterns would be imposed on different parts of a table. However, all requests associated to the same table are equally treated. DBAs' efforts to divide a table into multiple partitions [18], where each partition could get a different treatment, have been in an ad-hoc manner, without a guarantee of an effective performance optimization result. (3) Data placement policies that are configured according to the common access patterns of workloads have been largely static. Changes to the policies are avoided as much as possible, because data movement in a large granularity might interrupt user applications. Therefore, such static policies are difficult to adapt to the dynamic changes of I/O demands.

Monitoring-based storage management for databases can perform well when data accesses are stable in a long term, where certain regular access patterns can be identified via data access monitoring at runtime. Examples include general-purpose replacement algorithms in production systems: LRU, LIRS [12] and ARC [15], as well as recently proposed TAC [4] and Lazy-Cleaning [7]. However, monitoring-based management may not be effective under the following three conditions. First, monitoring-based methods need a period of ramp-up time to identify certain regular access patterns. For highly dynamic workloads and commonly found data with a short lifetime, such as temporary data, the ramp-up time may be too long to make a right and timely decision. Second, a recent study shows that data access monitoring methods would have difficulties to identify access patterns for concurrent streams on shared caching devices due to complex interferences [9]. In other words, concurrent accesses can cause unpredictable access patterns that may further reduce the accuracy of monitoring results. Third, certain information items are access-pattern irrelevant, such as content types and data lifetime [19], which are important for data placement decisions among heterogeneous storage devices. Monitoring-based approaches would not be able to identify such information. Furthermore, monitoring-based management needs additional computing and space support, which can be expensive to obtain a deep history of data accesses.

## 1.1 Outline of Our Solution: hStorage-DB

In order to address the limitations of DBA-based approach and particularly monitoring-based storage management, and to exploit the full capability of hybrid storage systems, we argue for a fundamentally different approach: making a direct communication channel between a DBMS and its underlying hybrid storage system. Our system framework is called heterogeneity-aware data management, or simplified as *hStorage-DB*.

We are motivated by the abundance of semantic information that is available from various DBMS components, such as the query optimizer and the execution engine, but has not been considered for database storage management. A DBMS storage manager is typically an interface to translate a DBMS data request into an I/O request. During the translation, all semantic information is stripped away, leaving only physical layout information of a request: logical block address, direction (read/write), size, and the actual data if it is a write. This in effect creates a *semantic gap* between DBMSs and storage systems.

In hStorage-DB, we bridge the semantic gap by making selected and important semantic information available to the storage manager, which can therefore classify requests into different types. With a set of predefined rules, each type is associated with a QoS policy that can be supported by the underlying storage system. At runtime, using the Differentiated Storage Services [16] protocol, the associated policy of a request is delivered to the storage system along with the request itself. Upon receiving a request, the storage system, first extracts the associated QoS policy, and then uses a proper mechanism to serve the request as required by the QoS policy.

As a case study, we have experimented with a hybrid storage system which is organized into a two-level hierarchy. Level one, consisting of SSDs, works as a cache for level two, consisting of HDDs. This storage system provides a set of caching priorities as QoS policies. Experiment results show the strong effectiveness of hStorage-DB.

Comparing with monitoring-based approaches [4, 7, 12, 15], hStorage-DB has the following unique advantages: (1) Under this framework, a storage system has the accurate information of how (and what) data will be accessed. This is especially important for highly dynamic query executions and concurrent workloads. (2) A storage system directly receives the QoS policy for each request, thus could quickly invoke the appropriate mechanism to serve. (3) Besides having the ability to pass access-pattern irrelevant semantic information, in hStroage-DB, storage management does not need special data structures required by various monitoring-based operations, thus incurs no additional computation and space overhead. (4) A DBMS can directly communicate with a storage system about the QoS policy for each request in an automatic mode, so that DBAs can be relieved from the burdens of storage complexity.

## 1.2 Critical Technical Issues

In order to turn our design of hStorage-DB into a reality, we must address the following technical issues.

**Associating a proper QoS policy to each request**: Semantic information does not directly link to proper QoS policies that can be understood by a storage system. Therefore, in order for a storage system to be able to serve a request with the correct mechanism, we need a method to accomplish the effective mapping from semantic information to QoS policies. However, a comprehensive solution must systematically consider multiple factors, including the diversity of query types, the complexity of various query plans, and the issues brought by concurrent query executions.

**Implementation of hStorage-DB**: Two challenges need to be addressed. (1) The QoS policy of each request eventually needs to be passed into the storage system. A DBMS usually communicates with storage through a block interface. However, current block interfaces do not allow passing anything other than the physical information of a request. (2) A hybrid storage system needs an effective mechanism to manage heterogeneous devices, so that data placement would match the unique functionalities and abilities of each device. Data also needs to be dynamically moved among devices to respond access pattern changes.

## 1.3 Our Contributions

This paper makes the following contributions. (1) We have identified a critical issue in the storage management of DBMSs, namely a semantic gap between the requirements of DBMS I/O requests and the supported services of heterogeneous storage services. Bridging this gap would significantly improve the performance of databases, particularly for complex OLAP queries and highly concurrent workloads, by addressing the limits of DBA-based and monitoring-based storage management approaches. We have designed hStorage-DB that restructures the storage management layer with semantic information to interface with the storage system. (2) We have implemented a system prototype of hStorage-DB that exploits the full capability of hybrid storage systems for database workloads by making informed, fine-grained and dynamic data block placement in a storage system. (3) We have evaluated the effectiveness of our prototype with a hybrid storage system, within which we use an SSD as a cache on top of hard disk drives. This storage system provides a set of caching priorities, that can be assigned to different types of requests. Performance of this system can be significantly improved by well utilizing the limited SSD capacity.

The rest of this paper is organized as follows. Section 2 outlines the architecture of hStorage-DB. Section 3 introduces QoS policies. Section 4 carries the core design of hStorage-DB by presenting a set of rules for assigning QoS policies to I/O requests. Section 5 overviews the Differentiated Storage Services protocol and its reference implementation (a hybrid storage system with caching priorities). Section 6 presents our experimental results. Section 7 discusses related work. Section 8 concludes this paper.

## 2. ARCHITECTURE OF hStorage-DB

Figure 1 shows the architecture of hStorage-DB. When the buffer pool manager sends a request to the storage manager, associated semantic information is also passed. We extend the storage manager with a "policy assignment table", which stores the rules to assign each request a proper QoS policy, according to its semantic information. The QoS policy is embedded into the original I/O request and delivered to the storage system through a block interface. We have implemented hStorage-DB by using the Differentiated Storage Services protocol from Intel Labs [16] to deliver a request and its associated policy to a hybrid storage system. Upon receiving a request, the storage system first extracts the policy, and invokes a mechanism to serve this request.

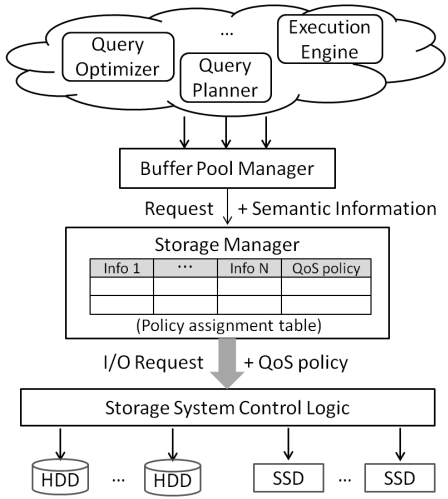

Figure 1: The architecture of hStorage-DB.

Our implementation of hStorage-DB is based on PostgreSQL 9.0.4. It mainly involves three issues: (1) We have instrumented the query optimizer and the execution engine to retrieve semantic information embedded in query plan trees and in buffer pool requests. (2) We have augmented the data structure of the buffer pool to store collected semantic information. The storage manager has also been augmented to incorporate the "policy assignment table". (3) Finally, since PostgreSQL is a multi-process DBMS, to deal with concurrency, a small region of the shared memory has been allocated for global data structures (Section 4.3) that need to be accessed by all processes.

The key to make hStorage-DB effective is associating each request with a proper QoS policy. In order to achieve our goal, we need to take the following 2 steps:

1. Understanding QoS policies and their storage implications;

2. Designing a method to determine an accurate mapping from request types to QoS policies.

In the following two sections, we will discuss the details.

## 3. QOS POLICIES

We will first discuss general QoS policies and then introduce the specific policies used in this paper.

### 3.1 Overview of QoS Policies

QoS policies provide a high-level service abstraction for a storage system. Through a set of well defined QoS policies, a storage system can effectively quantify its capabilities without exposing device-level details to users.

A QoS policy can either be performance related, such as latency or bandwidth requirements, or non-performance related, such as reliability requirements. All policies of a storage system are dependent on its hardware resources and organization of these resources. For example, if a storage system provides a reliability policy, then for an application running on such a storage system, when it issues write requests of important data, it can apply a reliability policy to these requests. Thus, when such a request is delivered, the storage system can automatically replicate received data to multiple devices.

On the one hand, QoS policies can isolate device-level complexity from applications, thus reducing the knowledge requirement on DBAs, and enabling heterogeneity-aware storage management within a DBMS. On the other hand, these policies determine the way in which a DBMS can manage requests. It is meaningless to apply a policy that cannot be understood by the storage system. Therefore, a different storage management module may be needed if a DBMS is ported to another storage system that provides a fundamentally different set of policies.

### 3.2 QoS Policies of a Hybrid Storage System

In this paper, we will demonstrate how to enable automatic storage management with a case study where the QoS policies are specified as a set of *caching priorities*.

The underlying system is a hybrid storage system prototype (detailed in Section 5) organized into a two-level hierarchy. The first level works as a cache for the second level. We use SSDs at the first level, and HDDs at the second level. In order to facilitate the decision making on cache management (which block should stay in cache, and what should not), its QoS policies are specified as a set of *caching priorities*, which can be defined as a 3-tuple:

$\{N, t, b\}$ , where $N > 0$, $0 \leq t \leq N$, and $0\% \leq b \leq 100\%$.

Parameter $N$ defines the total number of priorities, where a smaller number means a higher priority, i.e., a better chance to be cached.

Parameter $t$ is a threshold for "non-caching" priorities: blocks accessed by a request of a priority $\geq t$ would have no possibility of being cached. In this paper, we set $t = N - 1$. So, there are two non-caching priorities, $N - 1$ and $N$. We call priority $N - 1$ *"non-caching and non-eviction"*, and call $N$ *"non-caching and eviction"*.

There is a special priority, called *write buffer*, configured by parameter $b$. More details about these parameters will be discussed later.

For each incoming request, the storage system first extracts its associated QoS policy, and then adjusts the placement of all accessed blocks accordingly. For example, if a block is accessed by a request associated with a "high-priority", it will be fetched into cache if it is not already cached, depending on the relative priority of other blocks that are already in cache. Therefore in practice, the priority of a request is eventually transformed to the priority of all accessed data blocks. In the rest of paper, we will also use "priority of a block" without further explanation.

# 4. QOS POLICY FOR EACH REQUEST

In this section, we will present a set of rules that associate different I/O requests with appropriate QoS policies.

## 4.1 Request Types

A database I/O request has various semantic information. For the purpose of caching priorities, in this paper, we consider semantic information from the following categories.

**Content type**: We focus on three major content types: regular table, index and temporary data. Regular tables define the content of a database. They are the major consumers of database storage capacity. Indexes are used to speedup the accessing of regular tables. Temporary data, such as a hash table [8], would be generated during the execution of a query, and removed before the query is finished.

**Access pattern**: It refers to the behavior of an I/O request. It is determined by the query optimizer. A table may be either sequentially scanned or randomly accessed. An index is normally randomly accessed.

According to collected semantic information, we can classify requests into the following types: (1) *sequential requests*; (2) *random requests*; (3) *temporary data requests*; (4) *update requests*. The discussion of QoS policy mapping will be based on these types.

## 4.2 Policy Assignment in a Single Query

We will present five rules that are used to map each request type to a proper QoS policy which, in this case study, is a caching priority. For each request, the rules mainly consider two factors: 1) performance benefit if data is served from cache, and 2) data reuse possibility. These two factors determine if we should allocate cache space for a disk block, and if we decide to allocate, how long should we keep it in cache. In this subsection, we will consider priority assignment within the execution of a single query, and then discuss the issues brought by concurrent query executions in the next subsection.

### 4.2.1 Sequential Requests

In our storage system, the caching device is an SSD, and the lower-level uses HDDs, which can provide a comparable sequential access performance to that of SSDs. Thus, it is not beneficial to place sequentially accessed blocks in cache.

**RULE 1:** *All sequential requests will be assigned the "non-caching and non-eviction" priority.*

A request with the "non-caching and non-eviction" priority has two implications: (1) If the accessed data is not in cache, it will not be allocated in cache; (2) If the accessed data is already in cache, its priority, which is determined by a previous request, will not be affected by this request. In other words, requests with this priority do not affect the existing storage data layout.

### 4.2.2 Random Requests

Random requests may benefit from cache, but the eventual benefit is dependent on data reuse possibility. If a block is randomly accessed once but never randomly accessed again, we should not allocate cache space for it either. Our method to assign priorities for random requests is outlined in Rule 2.

**RULE 2:** *Random requests issued by operators at a lower-level of its query plan tree will be given a higher caching priority than those that are issued by operators at a higher-level of the query plan tree.*

This rule can be further explained with the following auxiliary descriptions.

**Level in a query plan tree:** For a multi-level query plan tree, we assume that the root is on the highest level; the leaf that has the longest distance from the root is on the lowest level, namely Level 0.

**Related operators:** This rule relates to random requests that are mostly issued by "index scan" operators. For such an operator, the requests to access a table and its corresponding index are all random.

**Blocking operators:** With a blocking operator, such as hash or sorting, operators at higher levels or its sibling operator cannot proceed unless it finishes. Therefore, the levels of affected operators will be recalculated as if this blocking operator is at Level 0.

**Priority range:** Note that there are totally N different priorities, but not all of them will be used for random requests. Instead, random requests are mapped to a consecutive priority range $[n_1, n_2]$, where $n_1 \leq n_2$. So, $n_1$ is the highest available priority for random requests; and $n_2$ is the lowest available priority.

**When multiple operators access the same table:** For some query plans, the same table may be randomly accessed by multiple operators. In this case, the priorities of all random requests to this table are determined by the operator at the lowest level of the query plan tree. If there is an operator that sequentially access the same table, the priority of this operator's requests is still determined by Rule 1.

Function (1) formalizes the process of calculating the priority of a random request, issued by an operator at Level $i$ of the query plan tree. Assume $l_{low}$ is the lowest level of all random access operators in the query plan tree, while $l_{high}$ is the highest level. And $L_{gap}$ represents this gap, where $L_{gap} = l_{high} - l_{low}$. Assume $C_{prio}$ is the size of the available priority range $[n_1, n_2]$, so $C_{prio} = n_2 - n_1$.

$$p(i) = \begin{cases} n_1 & \text{if } C_{prio} = 0 \\ n_1 & \text{if } L_{gap} = 0 \\ n_1 + i - l_{low} & \text{if } C_{prio} \geq L_{gap} \\ n_1 + \lfloor C_{prio} * \frac{i - l_{low}}{L_{gap}} \rfloor & \text{if } C_{prio} < L_{gap} \end{cases} \quad (1)$$

The last branch of this function describes the case when a tree has too many levels that there are not enough priorities to assign for each level. In this case, we can assign priorities according to the relative location of operators, and operators at neighboring levels may share the same priority.

Let us take the query plan tree in Figure 2 as an example. In this example, three tables are accessed: $t.a$, $t.b$ and $t.c$. We assume that the available priority range is [2,5]. Both operators that access $t.a$ are index scans. Since the lowest level of random access operators for $t.a$ is Level 0, all random requests to $t.a$ and its index would be assigned Priority 2. It also means that requests from the "index scan" operator at Level 1 are assigned the same priority: Priority 2.

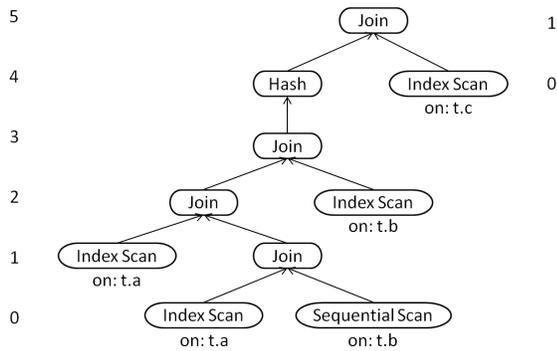

Figure 2: An example query plan tree. This tree has 6 levels. Root is on the highest level: Level 5. Due to the blocking operator "hash", the other two operators on Level 4 and 5 are re-calculated as on Level 0 and 1.

As to $t.b$, there are also two related operators. But according to Rule 1, all requests from the "sequential scan" operator (Level 0) are assigned the "non-caching and non-eviction" priority. Requests from the other operator (Level 2) that accesses $t.b$ are random. According to Function (1), requests from this operator should be assigned Priority 4.

For Table $t.c$, it is accessed by random requests from an "index scan" operator. However, due to the blocking operator "hash" on Level 4, the "index scan" operator is considered at Level 0 in priority recalculation, and thus all random requests to table $t.c$ would be assigned Priority 2.

### 4.2.3 Temporary Data Requests

Queries with certain operators may generate temporary data during execution. There are two phases associated with temporary data: *generation phase* and *consumption phase*. During generation phase, temporary data is created by a write stream. During consumption phase, temporary data is accessed by one or multiple read streams. In the end of consumption phase, the temporary data is deleted to free up disk space. Based on this observation, we should cache temporary data blocks once they are generated, and immediately evict them out of cache at the end of their lifetime.

**RULE 3:** *All read/write requests to temporary data are given the highest priority. The command to delete temporary data is assigned the "non-caching and eviction" priority.*

A request with the "non-caching and eviction" priority has two implications: (1) If the accessed data is not in cache, it will not be promoted into cache; (2) If the accessed data is already in cache, its priority will be changed to "non-caching and eviction", and can be evicted timely. Thus, requests with the "non-caching and eviction" priority only allow data to leave cache, instead of getting into cache.

Normally, if a DBMS is running with a file system, the file deletion command only results in metadata changes of the file system, without notifying the storage system about which specific blocks have become useless. This becomes a problem because temporary data may not be evicted promptly. And because of its priority, temporary data cannot be replaced by other data. Gradually, the cache will be filled with obsolete temporary data.

This issue can be addressed by the newly proposed TRIM command [1], which can inform the storage system of what LBA (logical block address) ranges have become useless due to file deletions, or other reasons. Supported file systems, such as EXT4, can automatically send TRIM commands once a file is deleted. For a legacy file system that does not support TRIM, we can use the following workaround to achieve the same effect: Before a temporary data file is deleted, we issue a series of read requests, with the "non-caching and eviction" priority, to scan the file from beginning to end. This will in effect tell the storage system that these blocks can be evicted immediately. This workaround incurs some overhead at an acceptable level, because the read requests are all sequential.

### 4.2.4 Update Requests

We allocate a small portion of the cache to buffer writes from applications, so that they do not access HDDs directly. With a write buffer, all written data will first be stored in the SSD cache, and flushed into the HDD asynchronously. Therefore, we apply the following rule for update requests:

**RULE 4:** *All update requests will be assigned the "write buffer" priority.*

There is a parameter $b$ that determines how much cache space is allocated as a write buffer. When the occupied space of data written by update requests exceeds $b$, all content in the write buffer is flushed into HDD. Note that the write buffer is not a dedicated space. Rather, it is a special priority that an update request can "win" cache space over requests of any other priority. For OLAP workloads in this paper, we set $b$ at 10%.

## 4.3 Concurrent Queries

When multiple queries are co-running, I/O requests accessing the same object might be assigned different priorities depending on which query they are from. To avoid such non-deterministic priority assignment, we apply the following rule for concurrent executions.

**RULE 5:**

1. *For sequential requests, temporary data requests and updates, the priority assignment still follows Rule 1, Rule 3 and Rule 4;*

2. *For random requests that access the same object (table or index) but for different queries, they are assigned the highest of all priorities, each of which is determined by Rule 3 and independently based on the query plan of each running query;*

To implement this rule, we store some global information for all queries to access: a hash table $H < oid, list >$, two variables $gl_{low}$ and $gl_{high}$.

1. The key "$oid$" stores an object ID that is either of a table or of an index.
2. The structure "$list$" for each $oid$ is a list.
3. Each element of $list$ is a 2-tuple $< level, count >$. It means that among all queries, there are totally $count$ operators accessing $oid$, and all of these operators are on Level $level$ in their own query plan tree. If some operators on different levels of a query plan tree are also accessing $oid$, we need another element to store this information.
4. Variable $gl_{low}$ (the global lowest level of all random operators) stores the minimum value of all $l_{low}$ according

to each query; similarly, $gl_{high}$ stores the maximum value of all $l_{high}$ according to each query.

All these data structures are updated upon the start and end of each query. To calculate the priority of a random request, with concurrency in consideration, we can still use Function (1), just changing $l_{low}$ into $gl_{low}$, and similarly $l_{high}$ to $gl_{high}$. and thus $L_{gap}$ would be $gl_{high} - gl_{low}$. Figure 3 describes this process.

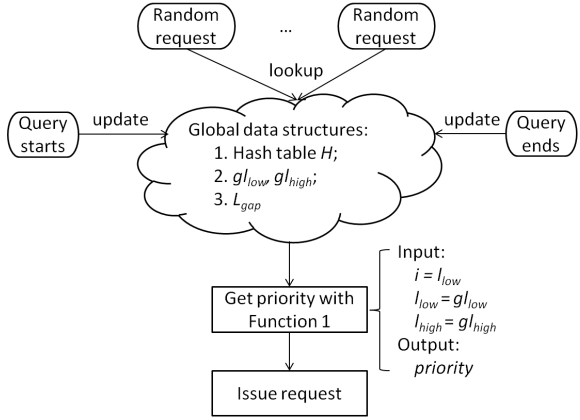

Figure 3: The process to calculate request priorities.

Table 1 summarizes all the rules hStorage-DB uses to assign caching priorities to requests.

| Request type | Priority | Rule |
|---|---|---|
| temporary data requests | 1 | Rule 3 |
| random requests | 2 … N-2 | Rules 2, 5 |
| sequential requests, | N-1 | Rule 1 |
| TRIM to temporary data | N | Rule 3 |
| updates | write buffer | Rule 4 |

Table 1: Rules to assign priorities.

## 5. STORAGE SYSTEM PROTOTYPE

The hybrid storage system we experimented with in this paper is a pre-release version of Intel's Open Storage Toolkit [11], which is organized into a two-level hierarchy. We use SSDs on the first level, working as a caching device for HDDs on the second level.

Using the Differentiated Storage Services protocol, an I/O request may not only contain physical information, but may also carry semantic information. This protocol provides backward compatibility with current block interfaces, such as SCSI, so that a classification-enabled DBMS can still run on top of a legacy storage system, while the semantic information of each request will simply be ignored.

### 5.1 Cache Management

As with any cache device, the most important data placement decisions are cache admission and eviction. They decide which data should be placed in cache and which data should be replaced. Within this storage system, both decision making processes are based on priorities. Therefore, they are called *selective allocation* and *selective eviction*.

- **Selective allocation:** Regarding the non-caching threshold $t$, only blocks with priorities $< t$ will be considered to cache. The final decision depends on the current cache capacity and relative priority of other in-cache blocks. For an incoming block, denoted as $N_{new}$, whose priority is $k$ and $k < t$, if the cache has additional space, it will be cached. Otherwise, if there exists a block, denoted as $N_{old}$, whose priority is $k'$ and $k' \geq k$, which means block $N_{old}$ has a lower priority than $N_{old}$. In this case, $N_{new}$ will also be cached, but after $N_{old}$ gets evicted (see below).

- **Selective eviction:** Eviction happens when cache needs to make room for new blocks. In order to determine which in-cache block should be evicted, the cache device first identifies blocks with Priority $k$, such that all other blocks in the cache have their priorities $< k$. Then among the blocks of Priority $k$, the "least-recently-used" one is selected to be evicted.

Cached blocks are organized into $N$ priority groups, where $N$ is the total number of priorities. Group $k, k \in [1, N]$, only contains blocks of Priority $k$. Each group is managed by the LRU (Least Recently Used) algorithm. Depending on the value of the "non-caching" threshold $t$, some low-priority groups may always be empty.

There are two types of blocks in cache: valid and invalid. Each valid block corresponds to a unique block within a level-two device, while invalid (or free) blocks do not correspond to any blocks in level-two devices. A valid block has two states: *clean* or *dirty*. A block is clean if there is an identical copy within a level-two device. Otherwise it is dirty. Based on selective allocation and selective eviction, the cache may perform one of the following six actions.

1. **Cache hit:** This action is taken when blocks accessed by an incoming request are already in cache. In this case, the caching device will directly communicate with OS. Based on the priority assigned to this request, there might be a follow-up action: "re-allocation". It will be explained later.

2. **Read allocation:** If the blocks accessed by an incoming read request are not in cache, but they are qualified for being cached, read allocation is involved. In this case, cache will first allocate enough space. Some in-cache blocks may be evicted if necessary. After that, new blocks will be read from level-two devices into cache, marked as clean, and then served to OS[1].

3. **Write allocation:** If the blocks contained in an incoming write request are not in cache, but they are qualified for being cached, write allocation is involved. After enough cache space is allocated, incoming blocks will be placed in cache, and marked as dirty. As soon as marking is done, the write request is returned. For both write allocation and read allocation, there might be data transmitted from cache to level-two devices, due to the eviction of dirty blocks.

---
[1]This is called synchronous read allocation, because data is placed into cache before the read request returns. Its opposite is asynchronous read allocation: blocks are served from level-two storage devices directly into OS, and placed into cache during idle time.

4. **Bypassing:** Bypassing is involved when blocks accessed by an incoming request are not in cache and are not qualified for being cached either. Since a storage-level cache is not necessarily on the critical path of data flow, for a read/write request with a low enough priority, its accessed blocks can be directly transmitted between OS and level-two devices, thus "bypassing" the cache.

5. **Re-allocation:** This action is taken when an accessed block is already in cache, but assigned a new priority. As described earlier, blocks in cache are organized into multiple priority groups. With a new priority, the block will be "removed" from its current group, and "inserted" into the corresponding new group.

6. **Eviction:** At times, a certain number of in-cache blocks (victims) should be evicted, to make room for new blocks. The selection of victims are described in the above "selection eviction". Once selected, victim blocks will be removed from their corresponding priority groups. For victim blocks that are also dirty, they will be written into level-two devices.

## 5.2 Metadata Management

In the storage system, blocks are managed by $N$ priority groups and a hash table. The total size of the metadata is proportional to the cache size.

The hash table is designed to facilitate the look-up of cached blocks. Each item in the hash table can be defined as $<lbn, V>$:

- $lbn$ is the logical block number. It is used as the hash key. If a logical block number is found in the hash table, the corresponding block is in cache; otherwise it is not cached.

- The value $V$ is itself a two-tuple $<pbn, prio>$. $pbn$ is a physical block address; it indicates where to find the block $lbn$ in the SSD cache. Since the cache device is invisible to the OS or applications, we cannot directly use $lbn$ to access a cached block. $prio$ stores the associated priority of the block, and it indicates which priority group this block belongs to.

All the data structures are stored in the main memory of the storage system.

## 6. PERFORMANCE EVALUATION

In this section, we will first present the effectiveness of hStorage-DB on accelerating executions of single queries. Then we will discuss concurrent workloads.

## 6.1 Experiment Setup

Our experiment platform consists of two machines, connected with a 10 Gb Ethernet link. One machine runs the DBMS, and the other is a storage server. Both are of the same configurations: 2 Intel Xeon E5354 processors, each having four 2.33GHz cores, 8 GB of main memory, Linux 2.6.34, two Seagate Cheetah 15.7K RPM 300 GB HDDs. The storage server has an additional Intel 320 Series 300 GB SSD to be our cache device. Key specifications of this SSD are shown in Table 2. Although a more high-end SSD would certainly improve cache performance, as we will find

| Sequential Read/Write | Random Read/Write |
|---|---|
| 270 MB/s / 205 MB/s | 39.5K IOPS / 23K IOPS |

Table 2: Performance specification of the cache device [10].

later, such an entry-level SSD could already demonstrate strong effectiveness of hStorage-DB.

On the storage server, one HDD runs the operating system; the other HDD and the SSD consist of the caching hierarchy as described in Section 5. The storage system is exported as a normal SCSI device to the DBMS server through iSCSI (Internet SCSI). On the DBMS server, all database data requests are sent to the storage server. except for transaction logs, which go to a dedicated local HDD.

We choose TPC-H [2] at a scale factor of 30 as our OLAP benchmark. With 9 indexes (shown in Table 3), the total dataset size is 46GB.

| 1 | lineitem (l_partkey); |
|---|---|
| 2 | lineitem (l_orderkey); |
| 3 | orders (o_orderkey); |
| 4 | partsupp (ps_partkey); |
| 5 | part (p_partkey); |
| 6 | customer (c_custkey); |
| 7 | supplier (s_suppkey); |
| 8 | region (r_regionkey); |
| 9 | nation (n_nationkey); |

Table 3: Indexes built for TPC-H.

## 6.2 Diversity of Request Types

Classification is meaningful only if a DBMS issues I/O requests of different types. In order to verify this assumption, for each query, we run it once, and count the number of I/O requests of each type, as well as the total number of disk blocks served for requests of each type.

As shown in Figure 4, we can observe requests of various types: *sequential requests*, *random requests* and *temporary data requests*.

## 6.3 Query Performance

For each query, we run it with the following four different storage configurations. (1) HDD-only; (2) LRU; (3) hStorage-DB; (4) SSD-only. HDD-only shows the baseline case when all I/O requests are served by a hard disk drive; SSD-only shows the ideal case when all I/O requests are served by an SSD; LRU emulates a classical approach when cache is managed by the LRU (least recently used) algorithm; In hStorage-DB, the storage system is managed under the framework as proposed in this paper. When we experiment with LRU and hStorage-DB, the SSD cache size is set to be 32GB, unless otherwise specified.

### 6.3.1 Sequential Requests

To demonstrate the ability of hStorage-DB to avoid unnecessary overhead for allocating cache space for low-locality data, we have experimented with Queries 1, 5, 11 and 19, whose executions are dominated by sequential requests, according to Figure 4. Test results are shown in Figure 5.

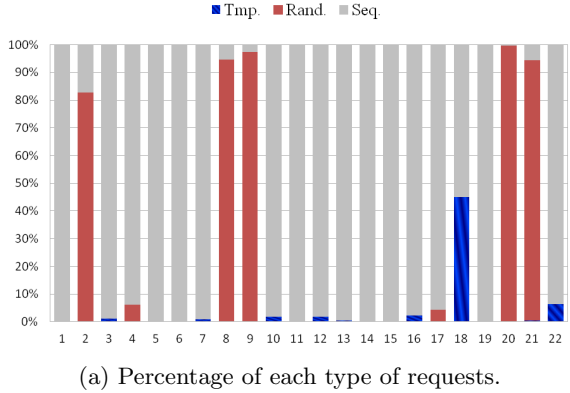
(a) Percentage of each type of requests.

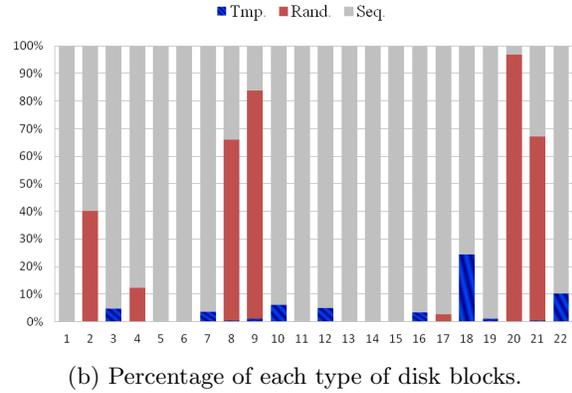
(b) Percentage of each type of disk blocks.

Figure 4: Diversity of IO requests in TPC-H queries. X-axis: Name of queries in TPC-H. Y-axis: Percentage of each type.

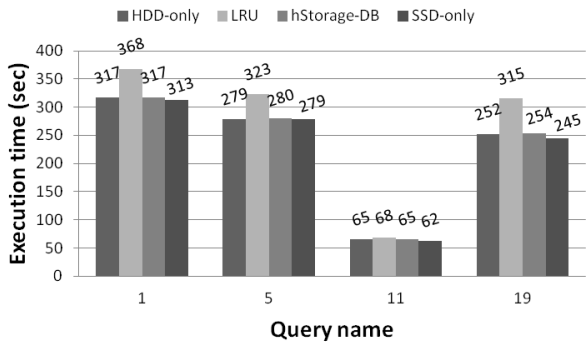

Figure 5: Execution times of queries dominated by sequential requests.

There are three observations from Figure 5: (1) The advantage of using SSD is not obvious for these queries. (2) If the cache is managed by LRU, which is not sensitive to sequential requests, the overhead can be significant. For example, compared with the baseline case, the execution time of LRU cache increased from 317 to 368 seconds for Q1, and from 252 to 315 seconds for Q19, resulting in a slowdown of 16% and 25% respectively. (3) Within the framework of hStorage-DB, sequential requests are associated with the "non-caching and non-eviction" priority, so they are not allocated in cache, and thus incurs almost no overhead.

|     | # of accessed blocks | # of hits | hit ratio |
| --- | --- | --- | --- |
| Q1  | 6,402,496 | 19,251 | 0.3% |
| Q5  | 8,149,376 | 17,694 | 0.2% |
| Q11 | 1,043,710 | 0 | 0% |
| Q19 | 6,646,328 | 16,798 | 0.3% |

Table 4: Cache statistics for sequential requests with LRU.

We have listed in Table 4 the number of accessed blocks and the number of cache hits for each query, when cache is managed by LRU. From this table, we can see that although caching data requested by sequential requests can bring cache hits for some queries, the hit ratio is negligible. Even the highest cache hit ratio is 0.3%, for Q1 and Q19.

### 6.3.2 Random Requests

In order to show the effectiveness of Rule 2 (Section 4.2.2), we have experimented with Q9 and Q21. Both have a significant amount of random requests, according to Figure 4. Results are plotted in Figure 6.

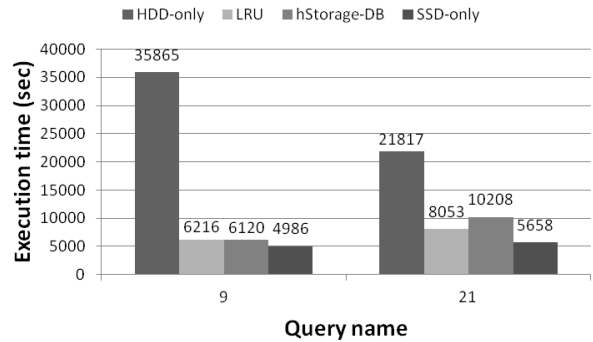

Figure 6: Execution times of queries dominated by random requests.

We have three observations from Figure 6. (1) The advantage of using SSD is obvious for the such queries. The performance of the ideal case (SSD-only) is far more superior than the baseline case (HDD-only). For Q9 and Q21, the speedup is 7.2 and 3.9 respectively. (2) Both queries have strong locality. LRU can effectively keep frequently accessed blocks in cache, through its monitoring of each block, and hStorage-DB achieves the same effect, through a different approach. (3) For Q21, hStorage-DB is slightly lower than LRU, which will be explained later.

To help better understand the performance numbers, Figure 7 shows the query plan tree of Q9. As we can see, there are two randomly accessed tables "supplier" and "orders". According to Rule 2, requests to "supplier" and its index are assigned Priority 2, and requests to "orders" and its index are assigned Priority 3.

Table 5 shows the cache statistics for requests of the two different priorities when executed by hStorage-DB. We can see that random requests of Q9 are served with a high cache hit ratio. LRU has a similar cache hit ratio, so we have omitted its numbers.

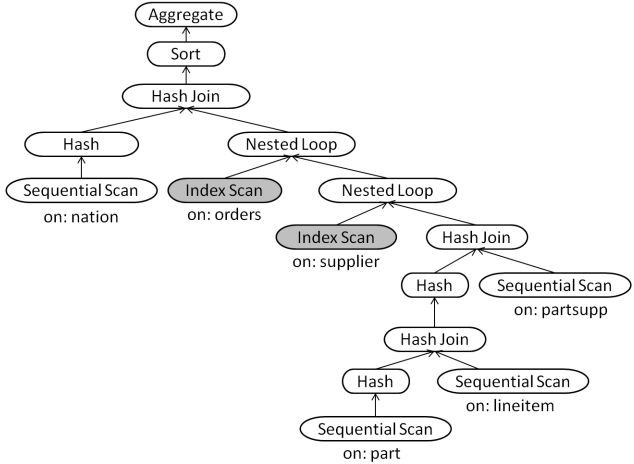

Figure 7: Query plan tree of Query 9.

|  | Priority 2 | Priority 3 |
|---|---|---|
| # of accessed blocks | 10,556,346 | 30,429,858 |
| Cache hits | 9,619,456 | 26,981,259 |
| Cache misses | 936,890 | 3,448,499 |
| hit ratio | 91% | 89% |

Table 5: Cache statistics for random requests of Query 9 with hStorage-DB.

Similarly, Figure 8 shows the query plan tree of Q21. According to this figure, tables "orders" and "lineitem" are randomly accessed. Based on Rule 2, requests to "orders" and its index are assigned Priority 2, and requests to "lineitem" and its index are assigned Priority 3.

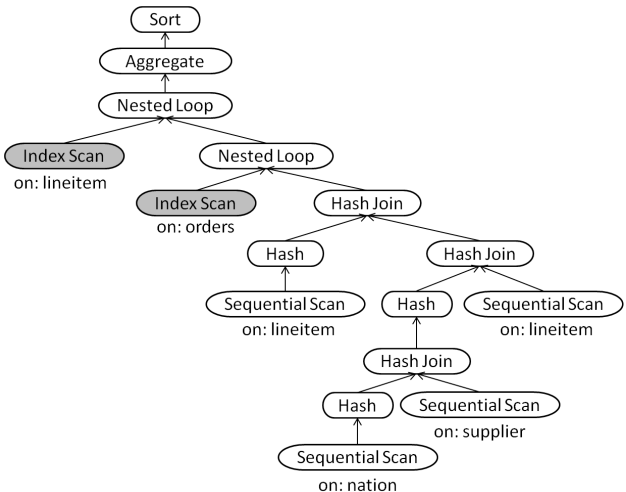

Figure 8: Query plan tree of Query 21.

| hStorage-DB | | | |
|---|---|---|---|
|  | Priority 2 | Priority 3 | Sequential |
| # of accessed blks | 18,353,605 | 11,591,715 | 12,816,956 |
| Cache hits | 16,585,399 | 7,366,930 | 147,656 |
| hit ratio | 90.3% | 40.1% | 0.1% |
| LRU | | | |
|  | Priority 2 | Priority 3 | Sequential |
| # of accessed blks | 18,211,959 | 10,876,511 | 12,816,959 |
| Cache hits | 16,430,097 | 8,954,023 | 6,524,852 |
| hit ratio | 90.2% | 82.3% | 50.9% |

Table 6: Cache hits/misses for Query 21.

Table 6 shows the cache statistics for requests of the two different priorities[2]. According to this table, both hStorage-DB and LRU can deliver a high cache hit ratio for Priority 2 requests, which are random requests to table "orders" and its index. But compared with hStorage-DB, LRU delivers a higher cache hit ratio for Priority 3 and sequential requests, which are all related to table "lineitem". As we can see from the query plan in Figure 8, "lineitem" is accessed by 3 operators, two "sequential scan" operators and one "index scan" operator. Therefore, they benefit from each other with LRU. In hStorage-DB, a block accessed by a sequential request would not be placed into cache, unless it is already cached. This is why, in terms of the execution time of Q21, hStorage-DB slightly underperforms LRU. However, equally treating sequential and random requests only benefits this case. We will show the disadvantages of being unable to discriminate sequential requests from other requests in Section 6.4.

### 6.3.3 Temporary Data Requests

In this section, we demonstrate the effectiveness of Rule 3 by experimenting with Q18, which generates a large number of temporary requests during its execution. Results are plotted in Figure 9.

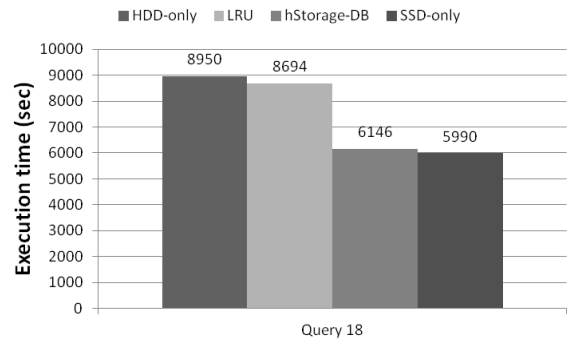

Figure 9: Execution time of Query 18.

This figure gives us the following three observations. (1) The advantage of using SSD is also obvious for this query, giving a speedup of 1.45 over the baseline case (HDD-only).

---
[2]In the lower half of the table, although we record statistics separately for requests of different priorities, all requests are managed through a single LRU stack.

(2) LRU also improves performance, because some temporary data blocks can be cached. But they are not kept long enough, so the speedup is not obvious. (3) hStorage-DB achieves a significant speedup because temporary data is cached until the end of its lifetime. The nature of "temporary data" can only be informed semantically.

As can be seen from the query plan (Figure 10), temporary data is generated by "hash" operators (in shaded boxes). We show the cache statistics in Table 7. Because writes of temporary data are all cache misses, we only consider temporary data reads. According to this table, LRU improves performance because some cached blocks are served from the cache, however, the cache hit ratio is only 1.8% for reads of temporary data. On the contrary, the hit ratio of temporary data reads is 100% with hStorage-DB.

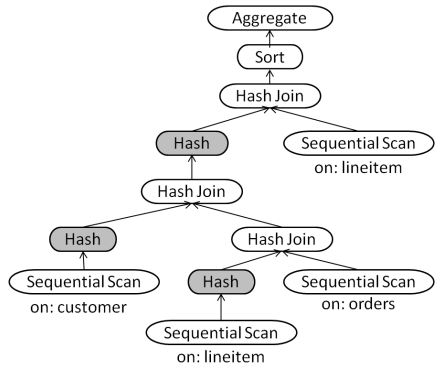

Figure 10: Query plan tree of Query 18.

| hStorage-DB | | |
|---|---|---|
| | Sequential | Temp. read |
| # of accessed blks | 19,409,504 | 5,374,440 |
| Cache hits | 0 | 5,374,440 |
| hit ratio | 0% | 100% |
| LRU | | |
| | Sequential | Temp. read |
| # of accessed blks | 19,409,358 | 5,374,486 |
| Cache hits | 64,552 | 96,741 |
| hit ratio | 0.3% | 1.8% |

Table 7: Cache hits of different blocks for Query 18.

#### 6.3.4 Running a Sequence of Queries

We have tested the performance of hStorage-DB with a stream of "randomly" ordered queries. We use the order of power test by the TPC-H specification [2]. We omit the results from a LRU-managed SSD cache, which has already been shown much less efficient than hStorage-DB.

Results are shown in Figure 11. In this figure, "RF1" is the update query at the beginning, and "RF2" is the update query in the end. For readability, the results of short queries and those of long queries are shown separately. According to the results, hStorage-DB shows clear performance improvements for most queries. In terms of total execution time of the sequence of queries, as shown in Table 8, hStorage-DB also improves significantly over the baseline case.

This experiment with a long sequence of queries took hStorage-DB more than 10 hours to finish, which shows the stability and practical workability of the hStorage-DB solution. Different from running each query independently, the success of this experiment involves the issues of data reuse and cache space utilization during a query stream. Particularly, useless cached data left from a previous query need to be effectively evicted from cache. Experiment results have shown that in the framework of hStorage-DB, (1) temporary data can be evicted promptly, by requests with the "non-caching and eviction" priority; and (2) data randomly accessed by a previous query can be effectively evicted by random requests of the next query, if it will not used again.

| HDD-only | hStorage-DB | SSD-only |
|---|---|---|
| 86,009 | 39,132 | 23,953 |

Table 8: Total execution time (seconds) of the sequence.

### 6.4 Concurrency

We have run a throughput test by the TPC-H specification [2]. In this test, we set the scale factor at 10, and the total dataset size was 16GB. We used 2GB main memory and 4GB SSD cache. During the test, we used 3 query streams and 1 update stream. Table 9 shows the overall results.

| HDD-only | LRU | hStorage-DB | SSD-only |
|---|---|---|---|
| 13 | 28 | 43 | 114 |

Table 9: TPC-H throughput results.

According to this table, for throughput test, hStorage-DB has a $3.3X$ speedup over the baseline case, and a $1.5X$ speedup over the performance of LRU. We can see that these speedups are much larger than those observed in previous experiments for single queries. To explain this, we look into the execution time of different types of queries in throughput test and single query test.

To clearly show the benefits of hStorage-DB, we closely study two queries: Q9 and Q18. We have confirmed that at a scale factor of 10, Q9 also has a significant number of random requests, while Q18 also has many temporary data requests. Figure 12a shows their performance numbers when running independently. Figure 12b shows the average execution time of the two queries in the throughput test.

For Q9, hStorage-DB can still give a performance result that is close to the ideal case (SSD only). This is because the data randomly accessed by this query has been successfully protected from other cache-polluting data during its execution. In comparison, even though the performance of LRU is close to that of hStorage-DB when running Q9 independently, in the case of concurrent workloads, it is 2.8 times lower than hStorage-DB.

Q18 is another example, which has temporary data during execution. When running Q18 independently, LRU is only 1.2 times slower than hStorage-DB, but in the case of concurrent workloads, it becomes 1.85 times slower.

### 6.5 Summary of Experiment Results

Our experiments have shown the effectiveness of hStorage-DB in three unique ways: (1) For workloads whose access patterns cannot be effectively detected by monitoring-based methods, hStorage-DB can directly pass critical semantic

1085

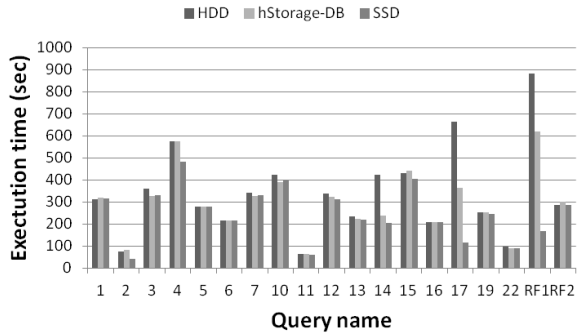
(a) Short queries

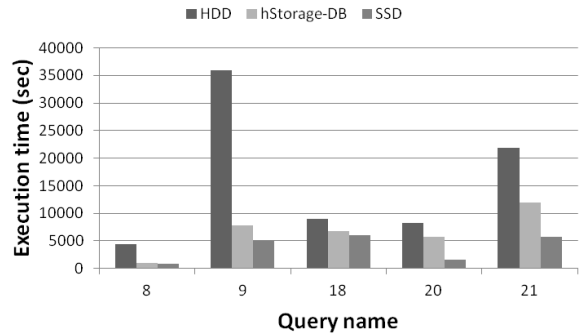
(b) Long queries

Figure 11: Execution times of queries when packed into one query stream.

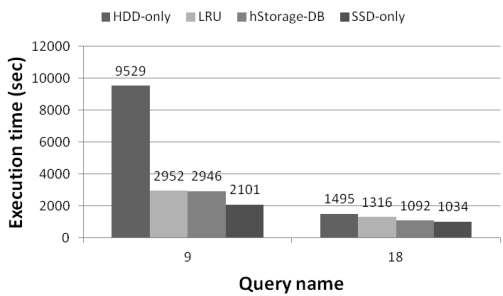
(a) Standalone execution times.

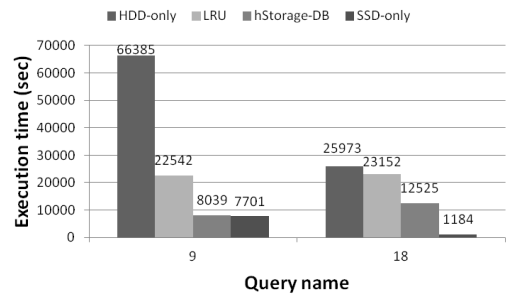
(b) Average execution times in throughput test.

Figure 12: Comparison of the execution times for Q9 and Q18.

information to the storage system for effective data placement decisions (results presented in Section 6.3.1). In addition, for workloads whose access patterns can be effectively detected by monitoring-based methods, hStorage-DB could achieve comparable performance (results presented in Section 6.3.2). (2) The hStorage-DB framework can well utilize access-pattern irrelevant semantic information items, e.g., content type and lifetime of temporary data. Such information items play important roles for storage management (results presented in Section 6.3.3). (3) Within an environment of concurrent query executions, hStorage-DB shows its strong advantages over monitoring-based methods by accurately recognizing the relative importance of different data blocks, so that cache pollution could be effectively prevented and important data is cached as long as necessary (results presented in Section 6.4).

## 7. OTHER RELATED WORK

There have been several different approaches to managing storage data in databases, each of which has unique merits and limits. The most classical approach is to apply a replacement mechanism to keep or evict data at different levels of the storage hierarchy. Representative replacement algorithms include LIRS [12] that is used in MySQL and other data processing systems, and ARC [15] that is used in IBM storage systems and ZFS file system. The advantage of this approach is that it is independent of workloads and underlying systems. The nature of general-purpose enables this approach to be easily deployed in a large scope of data management systems. However, the two major disadvantages are (1) this approach would not take advantage of domain knowledge even it is available; and (2) it requires a period of monitoring time before determining access patterns for making replacement decisions.

Another approach is more database storage specific. There are two recent and representative papers focusing on SSD management in this category. In [4], authors propose an SSD buffer pool admission management by monitoring data block accesses to disks and distinguishing blocks into warm regions and cold regions. A temperature-based admission decision to the SSD is made based on the monitoring results: admitting the warm region data and randomly accessed data to the SSD, and making the cold region data stay in hard disks. In [7], authors propose three admission mechanisms to SSDs after the data is evicted from memory buffer pool. The three alternative designs include (1) clean write (CW) that never writes the dirty pages to the SSD; (2) dual-write (DW) that writes dirty pages to both the SSD and hard disks; and (3) lazy-cleaning (LC) that writes dirty pages to the SSD first, and lazily copies the dirty pages to hard disks. Although specific to the database domain, this approach has several limitations that are addressed by hStorage-DB.

Compared the aforesaid prior work, hStorage-DB leverages database semantic information to make effective data placement decisions in storage systems. Different from application hints, which can be interpreted by a storage system in different ways or simply ignored [14], semantic information in hStorage-DB requests a storage system to make data placement decisions. In particular, hStorage-DB has the following unique advantages.

First, rich semantic information in a DBMS can be effectively used for data placement decisions among diverse storage devices, as have been shown in our experiments. Existing approaches have employed various block-level detection and monitoring methods but cannot directly exploit such semantic information that is already available in a DBMS.

Second, some semantic information that plays an important role in storage management cannot be detected. Take temporary data as an example. The hStorage-DB performs effective storage management by (1) caching temporary data during its lifetime, and (2) immediately evicting temporary data out of cache at the end of its lifetime. In TAC [4], temporary data writes would directly go to the HDD, instead of the SSD cache, because such data are newly generated, and are not likely to have a "dirty" version in the cache that need to be updated. In the three alternatives from [7], only DW and LC are able to cache generated temporary data. However, in the end of lifetime, temporary data cannot be immediately evicted out of cache to free up space for other useful data.

Furthermore, some semantic information, such as access patterns, may be detected, but with a considerable cost. Within hStorage-DB, such information is utilized with no extra overhead. For small queries, the execution may be finished before its access pattern could be identified. One example is related to sequential requests. In [7], special information from SQL Server is used to prevent special sequential request from being cached in SSD cache. In contrast, hStorage-DB attempts to systematically collect critical semantic information in a large scope.

## 8. CONCLUSION AND FUTURE WORK

We have identified two sets of problems related to DBA-based and monitoring-based storage management approaches for database systems with heterogeneous storage devices. These problems can be addressed by filling in the semantic gap between a DBMS and its storage system. We have proposed and implemented hStorage-DB to achieve this goal. Instead of relying on the storage system to detect the best way to serve each I/O request, hStorage-DB takes a top-down approach in three steps: classification for each request, a mapping from each request type to a proper QoS policy, and invoking the right mechanism that is supported by the storage system to serve the request properly. In the framework of hStorage, QoS policies provide a high level abstraction of the services supported by a storage system. This system enables block-granularity and dynamic data placement. Our experiment results have shown the strong effectiveness of hStorage-DB. We are currently extending hStorage-DB for OLTP workloads. We will also consider semantic information from database applications.

## 9. ACKNOWLEDGMENTS

We sincerely thank the anonymous reviewers for their constructive comments. The work was supported in part by the National Science Foundation under grants of CCF-0913050 and CNS-1162165.